\RequirePackage{fix-cm}
\documentclass[twocolumn,epjc3]{svjour3}

\smartqed

\input{pkgs}

\title{Challenges for dark matter direct search with SiPMs}

\newcommand{\AQLNGS}{INFN Laboratori Nazionali del Gran Sasso, Assergi (AQ) 67100, Italy}

\author{
  Alessandro~Razeto\thanksref{LNGS}\thanksref{email1}
  \and Nicola~Rossi\thanksref{LNGS}\thanksref{email2}
}

\institute{
  \AQLNGS \label{LNGS}
}

\thankstext{email1}{E-mail address: alessandro.razeto [at] lngs.infn.it}
\thankstext{email2}{E-mail address: nicola.rossi [at] lngs.infn.it}

\journalname{} 
\date{}

\begin{document}
\maketitle

\begin{abstract}
Liquid xenon and liquid argon detectors are leading the direct dark matter search and are expected to be the candidate technology for the forthcoming generation of ultra-sensitive large-mass detectors. At prese- nt, the scintillation light detection in those experiments is based on ultra-pure low-noise photo-multipliers. To overcome the issues in terms of the extreme radio-purity, costs, and technological feasibility of the future dark matter experiments, the novel SiPM-based photo-detec- tor modules look promising candidates, capable of replacing the present light detection technology. However, the intrinsic features of SiPMs may limit the present expectations. In particular, interfering phenomena, especially related to the optical correlated noise, can degrade the energy and pulse shape resolutions. As a consequence, the projected sensitivity of the future detectors has to be reconsidered accordingly.

\keywords{Noble gas detector \and TPC  \and light yield \and dark matter \and SiPM \and liquid argon \and liquid xenon}
\end{abstract}

\section*{Introduction}

Indirect astrophysics observations from galactic to large structure scales require the presence of a non-luminous and hidden mass, commonly named \emph{dark matter}, accounting for about 85\% of the matter content of the Universe (for a review, see e.g.~\cite{bib:PDG}). Although the nature of this missing ingredient is still unknown, many particle candidates have been proposed over the recent decades, with masses ranging from fraction of eV/c$^2$ to the Grand Unification mass scale. Weakly interacting massive particles (WIMP), with mass in the range from 1 GeV/c$^2$ up to 1 TeV/c$^2$~\cite{bib:WIMP1, bib:WIMP2}, have been considered promising candidates, as originating from natural high energy extensions of the Standard Model of particle physics. However, the lack of observation in many direct and indirect searches is recently making questionable their theoretical foundations~\cite{bib:Bertone}.

Since the beginning of the new millennium, the direct dark matter search has been attracting a lot of interest in the astroparticle physics community, with a substantial effort in terms of engagement and funding. This enormous enterprise seems not to pay back yet. Since the majority of the experiments searching for WIMPs are providing null results, a quest is ongoing for the deployment of the ultimate hundred-ton-scale targets. The latter effort requires experimental technology easily scalable, while preserving a very low level of background, high detection efficiency, and stable behavior over time exposures of tens of years. 

One of the most successful and scalable detection technologies is based on liquid noble gas, in particular on xenon and argon. Both targets show clear \emph{pros and cons} and, at the moment, the scientific community is split down the middle, and there is no clear choice for the ultimate experiment capable of reaching the so-called \emph{neutrino floor} limit in the forthcoming decades~\cite{argo, bib:DARWIN}. 
The experiments based on xenon double-phase time projection chamber (TPC), as LZ~\cite{bib:LZ}, XE- NON-1T~\cite{bib:XENON-1T}, PANDAX-4T~\cite{bib:PandaX-4T} reported independently the strongest bound on the spin-independent  WIMP-nucleon interaction (SI). 
DEAP-3600~\cite{deap3600}, a single phase liquid argon detector, reported an independent, even if lower, limit for the same interaction.
Finally, DarkSide-50, using ultra-pure underground argon in a double-phase TPC, has set a corresponding robust limit especially for the low dark matter masses in the region down to a few GeV/c$^2$~\cite{ds532, ds50low}. 

Recently, LZ has reported the strongest SI limit \cite{lzfirst} and XENON-nT has shown a very low background in the first physical data-taking of its multi-ton target~\cite{xenonnt-el}.
Meanwhile, DarkSide-20k \cite{ds20k}, that will exploit about 20 ton (fiducial mass) of ultra-pure underground argon in a gigantic double-phase TPC, is under construction and should start the data-taking by the end of the present decade. 
The argon and xenon scientific communities are already designing the next iteration of experiments: the xenon community is converging on a future detector, called DARWIN, based on 50~ton of target~\cite{bib:DARWIN}, while the argon community is planning a future detector, called ARGO-300, based on a target of 300~ton of ultra-pure argon naturally depleted in \isotope[39]{Ar}~\cite{argo}. 

At present, the scintillation light of the aforementioned experiments is detected by ultra-pure low-noise photo-multiplier tubes (PMTs) \cite{bib:LZ, bib:XENON-nT, bib:PandaX-4T, ds532, deap3600}. 
To overcome possible issues in terms of the extreme radio-purity, costs, and technological feasibility, many projects are planning to replace the PMTs with novel photo-detectors. In particular, the DarkSide-20k experiment, supposed to reach 200 ton$\cdot$y exposure, will install about 22 m$^2$ of silicon photo-multipliers (SiPM) arrays~\cite{ds20k, tile, MB4}. 

In this work, extensive studies on the SiPM performances prove that the intrinsic features of the SiPM units, in addition to the well known internal correlated noise,  may limit the present expectations. In particular, interfering phenomena related to the optical correlated noise, i.e. the light emission of fired SiPMs triggering nearest SiPMs, produces an irreducible degradation of the energy and pulse shape resolutions. As a consequence, the projected sensitivity of the future detectors, especially based on liquid argon, has to be reconsidered accordingly. 

The articles is sketched according to the following structure: in Sec. \ref{sec:noble} the main argon- and xenon-based characteristics concerning the light detection are described; in Sec. \ref{sec:sipm} the main feature of SiPM-based photo-detector module at cryogenic temperature is reviewed, and the noise classification and the impact on the event reconstruction is largely discussed; In Sec.  \ref{sec:tmc} a toy Monte Carlo simulation for a multi-ton liquid argon TPC
is performed, and the impact of the correlated noise in the final dark matter analysis is outlined. Finally, in Sec. \ref{sec:sensitivity} the implication of the irreducible correlated noise on the dark matter sensitivity plots is discussed.

\section{Liquid noble gas detectors}
\label{sec:noble}

Noble liquid detectors can exploit either single~\cite{deap3600} or double-phase~\cite{bib:TPB, ds20k, bib:XENON-nT, bib:LZ, bib:PandaX-4T}. The first type has a simpler design and shows fewer technological issues. In those detectors, like DEAP-3600~\cite{deap3600}, the liquid noble gas fills up a spherical vessel whose surface is instrumented with PMTs to detect the scintillation light. 
The reconstructed position is not that accurate and it is in principle difficult to reject the pile-up events. Whereas, the second type shows a more complex design, but features many advantages in terms of event reconstruction. A double-phase TPC is typically a cylindrical-shaped vessel filled with a noble liquid. On top of the liquid free surface there is a thin gas pocket, separated by a conductive grid. On the bottom and on the top of the cylinder there are a cathode and an anode respectively, able to set the so-called  \emph{drift} and \emph{extraction} fields. For each ionizing particle hitting the detector, two signals are produced: the first scintillation signal is originated in the liquid by the primary interaction and is typically called S1, whereas the second signal (S2) is produced by electro-luminescence of the gas pocket when ionization electrons, pulled upwards by the drift fields, are accelerated in the gas pocket by the extraction field. The double signal allows to increase the spacial resolution and exploit the S2/S1 ratio pulse shape discrimination between electron recoils (ER) and nuclear recoils (NR).

Among the available noble gases in nature, at present only argon and xenon have shown a reasonable feasibility in terms of costs and performances. Both technologies have proven to be easily scalable to multi-ton scales, with no insuperable obstacles and with reasonable economical effort. However, the two gases have very different physical properties being, to some extent, complementary. This is especially true as well for the design of PMT- and SiPM-based photo-detectors.

The use of the PMTs in liquid argon has in fact experienced a very critical performance in terms of electronics stability, do to the very low cryogenic temperature of 87 K. In addition, the 128 nm scintillation light does not match the PMT photo-cathodic sensitive window, and then a wavelength shifter coated reflector is therefore needed (typically made of by Tetraphenyl butadiene (TPB) at 420 nm \cite{bib:TPB}). 
On the contrary, in liquid argon the use of SiPMs is encouraged by the low dark rate at 87 K and by the easy detection of visible shifted scintillation light that matches the high particle detection efficiency (PDE) of commercial SiPMs~\cite{nuv-hd-cryo, zappala}.

For the xenon-based detector, instead,  the use of PMTs is much handier, since they operate steadily at the xenon cryogenic temperature (165 K) and are highly sensitive to the scintillation light at 178 nm without the need of a wavelength shift. Concerning the use of SiPMs in xenon, in principle photo-cathodic surfaces with a reasonable PDE to vacuum ultra violet (VUV) light are already available. However, the higher cryogenic temperature could non be sufficiently low to make the dark rate down to an acceptable threshold.

\section{SiPM-based photo-detectors}
\label{sec:sipm}

The SiPMs are solid-state single-photon-sensitive devices based on single-photon avalanche diode (SPAD) micro-cells on a silicon substrates~\cite{buzhan}. The dimension of each single SPAD ranges between \qtyrange[range-phrase = \ \text{and}\ ]{10}{100}{\micro\meter}.
Each SPAD operates in Geiger mode, coupled with the others by a quenching circuit. 
For analog SiPMs, the signal of the micro-cells is summed in parallel by appropriate quenching resistors: this results in a dynamic range spanning from 1 to thousands photo-electrons (for mm$^2$) with an intrinsic photon counting resolution exceeding few percent~\cite{cryo-pre}.
SiPMs produce a signal proportional to the bias exceeding the break-down voltage ($V_B$, typically between \qtyrange[range-phrase = \ \text{and}\ ]{20}{50}{V}) with over-voltages (OV) in the range 2--10 V, well below $V_B$~\cite{piemonte}.

Since first measurements at cryogenic temperature ~\cite{COLLAZUOL2011389}, the Fondazione Bruno Kessler (FBK) intensified the effort for the development of SiPMs for cryogenic particle detectors. 
The current NUV-HD-Cryo family is capable of stable operation in liquid nitrogen/argon at over-voltages in excess of \qty{10}{V}~\cite{2pac}. These devices are optimized for the detection of near ultra-violet (NUV) and blue light with a peak photon detection efficiency close to \qty{60}{\percent}~\cite{nuv-hd-cryo}, measured at room temperature. The primary dark rate (DCR) is lower than few counts per square centimeter per second at \qty{77}{\kelvin}~\cite{cryo-nuv}.

\begin{figure}[tb]
\centering
\includegraphics[width=\columnwidth]{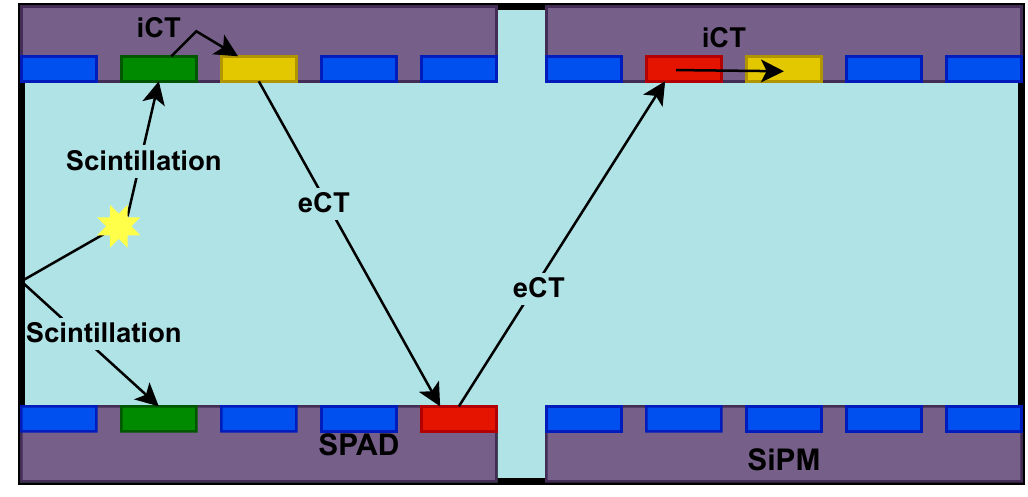}
\caption{Photon detection in a high coverage detector with SiPMs: scintillation photo-electron may be followed by a cascade of internal and external cross-talk. The SiPMs are drawn in gray with blue SPADs that become green/red/yellow when triggered respectively by a primary photon, by eCT and by iCT (see text). All surfaces of the chamber are highly reflective; the scintillation medium is shown in cyan. For simplicity after-pulsing is not shown.}
\label{fig:sketch}
\end{figure}

Beyond DCR, SiPMs exhibit correlated noises: after-pulsing (AP) occurs when, during an avalanche, a carrier is temporarily trapped by impurities in the medium. When released, a second avalanche is generated. If the detrapping happens before the SPAD is fully recharged, the after-pulse will have a charge lower than the single photo-electron. 
\emph{Optical cross-talk} (oCT) is triggered by photons generated as secondary process during the avalanche. These photons can interact with a neighbour micro-cell and generate \emph{internal cross-talk} (iCT), or escape the silicon bulk and generate \emph{external cross-talk} (eCT) in the nearest SiPMs.
Figure~\ref{fig:sketch} sketches the optical-crosstalk cascade in case of high photo-detection coverage. Consider that in TPCs, due to the high reflectivity of the chosen materials,  the collection efficiency for photons can typically exceed \qty{90}{\percent}~\cite{star}.

Finally, the final signal is inevitably affected by electronic noise (EN) accounting either for the ergodic processes (Johnson–Nyquist, Schottky and 1/f noises) and coupled noise, both amplified and shaped by the read-out electronics.

At Laboratori Nazionali del Gran Sasso (LNGS-INFN), large integrated photo-detectors for cryogenic applications have been developed~\cite{cryo-pre, tile}: it is now possible to aggregate the signals of \num{96} SiPMs into a single analog output covering a surface of \qty{100}{\square\cm}~\cite{MB4}. As a result, it is now possible to plan large dark matter experiments using SiPMs for light detection~\cite{ds20k}. 

\subsection{Optical cross-talk models}

The cross-talk cascade process leads to an information-less multiplication of the number photo-electron, defined here as detection \emph{noise gain} $\mathcal{G}_\textrm{oCT}$ (not be confused with the charge gain of SiPM avalanche). In other words, the apparent light yield of a particle detector can be significantly larger than the real one, since for each real photo-electron (associated to the scintillation photon), a number $\mathcal{G}_\textrm{oCT}\, (>1)$ are in average actually detected.

As described in~\cite{vinogradov-coupound}, iCT follows a recursive process in which the primary avalanche is followed by secondaries, which in turn can produce other avalanches. The process is theoretically limited to the very large number of micro-cells in the detector (between millions and billions), but for mathematical convenience, the geometric progression is approximated with a series.
Let $\lambda_\textrm{iCT}$ be the mean number of secondaries for each primary avalanche ($\lambda < 1$), the noisy detection gain converges to $\mathcal{G}_\textrm{iCT}=1/(1-\lambda_\textrm{iCT})$ with an excess noise factor $\mathcal{E}_\textrm{iCT} \simeq 1 + \lambda_\textrm{iCT}$~\cite{vinogradov-coupound}. 
The fluctuations in the cascade have strong impact on the resolution of a particle detector using SiPMs: it is possible to introduce a generalized Fano factor ($\mathcal{F}_\textrm{iCT} = \mathcal{G}_\textrm{iCT} \,\mathcal{E}_\textrm{iCT}$), that directly correlate the number of detected photons with its variance.

\begin{figure}[tb]
\centering
\includegraphics[width=0.6\columnwidth]{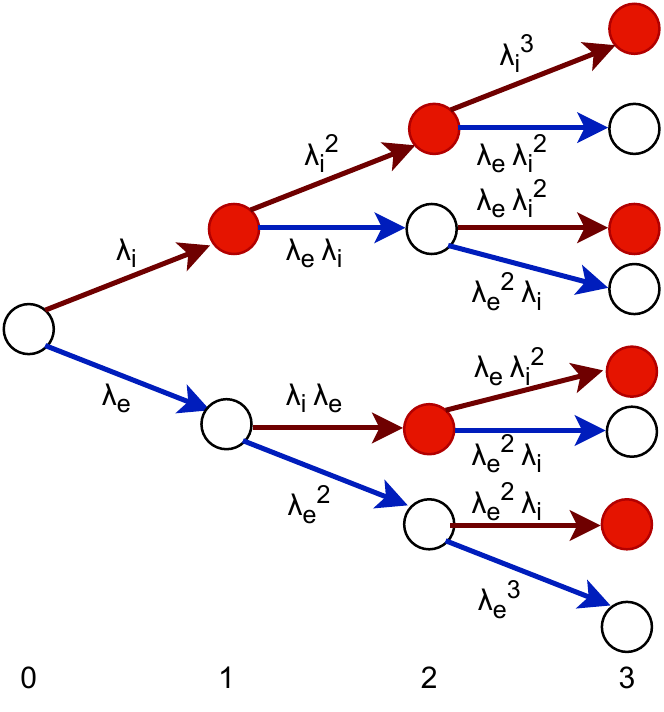}
\caption{Optical cross-talk cascade in presence of both internal and external processes. For level $n$-th of recursion the average number of avalanches is $N^n=\left(\lambda_\textrm{iCT} + \lambda_\textrm{eCT}\right)^n$ (for brevity the CT pedix has been removed in the figure) while the number of iCT only photo-electrons (red cirles) is $N^n_\textrm{iCT}= N^{n-1}\, \lambda_\textrm{iCT}$ (n>0). 
The sum over i converge to $\mathcal{G}_\textrm{oCT}$, defined in Equation~\ref{eqn:gain:oct}, and $N_\textrm{iCT}=\mathcal{G}_\textrm{oCT}\,\lambda_\textrm{iCT}$.}
\label{fig:gain}
\end{figure}

The modeling of the iCT is further complicated by neighbours effect, for which already triggered micro-cells reduce the acceptance for new photons: therefore both the ``Branching Poisson'' and the ``Geometric Chain'' models introduced in  ~\cite{ct-modelling} result inaccurate.\\

For a simplified symmetric detector ( where all the SiPMs are identical and the optical acceptances for external cross-talk is uniform), using positive feed-back theory, it is possible to write
\begin{equation}
    N_G^\textrm{pe} =     N_N^\textrm{pe} + \lambda_\textrm{iCT}\,     N_G^\textrm{pe} + \lambda_\textrm{eCT}\, N_G^\textrm{pe}
\label{eqn:gain:impl}
\end{equation}
where $N_G^\textrm{pe}$ and $N_N^\textrm{pe}$ represent the detected (gross) and scintillation (net) number of photo-electrons.
Equation~\ref{eqn:gain:impl} implies only the additive and independence properties of iCT and eCT photons. With these assumptions, the oCT gain becomes by extension:
\begin{equation} 
  \mathcal{G}_\textrm{oCT}=\frac{1}{1-\lambda_\textrm{iCT}-\lambda_\textrm{eCT}}
  \label{eqn:gain:oct}
\end{equation}
that is valid for $\lambda_\textrm{oCT} =  \lambda_\textrm{iCT}+\lambda_\textrm{eCT} < 1$, otherwise the system diverges. 
It may be interesting to know how many eCT photo-electrons are generated in average for each physical scintillation photo-electrons, discarding the contribution of iCT. The solution is simply
\begin{align}
\begin{split}
    N^\textrm{pe}_\textrm{iCT} &= \mathcal{G}_\textrm{oCT}\,\lambda_{\rm iCT}\, N_N^\textrm{pe}\\
    N^\textrm{pe}_\textrm{eCT} &= \mathcal{G}_\textrm{oCT}\,\lambda_{\rm eCT}\, N_N^\textrm{pe}
    \label{eqn:tartaglia}
\end{split}
\end{align}
as the iCTs and eCTs probability are in fact additive. Figure~\ref{fig:gain} reports a visual demonstration of these formulas for the geometric chain hypothesis. However, these are valid as well in the branching poisson model, as  application of the multinomial theorem.

\subsection{After-pulsing}
The AP adds another positive feedback contribution to the process, thus increasing the overall noise gain. 
Modifying Eq.~\ref{eqn:gain:oct} to include for the after-pulsing is not trivial because the charge gain of the avalanches depends on the delay from the primary photo-electron.
At leading order, one can write
\begin{equation}
  \mathcal{G}_\textrm{oCT+AP}=\frac{1}{1-\lambda_\textrm{iCT}-\lambda_\textrm{eCT}-\lambda^*_\textrm{AP}}
  \label{eqn:gain:octap}
\end{equation}
where $\lambda^*_\textrm{AP}$ is an effective parameters.

The AP probability ($P_\textrm{AP}^T$) in literature is measured above a fixed signal amplitude, typically T=\qty{50}{\percent}. 
The amplitude of the after-pulsing is proportional to the status of the recharge of the fired micro-cell. The signal half life for the photo-detectors in consideration is about $T^{(\nicefrac{1}{2})}=$\qty{100}{\nano\second}~\cite{MB4}. This means that for after-pulsing closer than $T^{(\nicefrac{1}{2})}$ the charge of the avalanche is more than halved. 

The probability of emitting oCT photons is proportional to the charge of the avalanche, therefore after-pulsing with low amplitude will not contribute significantly to $\mathcal{G}_\textrm{oCT+AP}$.  
For the NUV-HD-Cryo SiPM at the highest over-voltage $P_\textrm{AP}^{(\nicefrac{1}{2})}$  is less than one tenth of the combined oCT lambdas (see Tab.~\ref{tab:tmc}), therefore (at leading order) it can be assumed $\lambda^*_\textrm{AP}=P_\textrm{AP}^{(\nicefrac{1}{2})}$.

\begin{table}[tb]
\centering
\begin{tabular}{|p{17mm}|c|c|c|c|c|}
\hline 
Parameter & \qty{5}{OV} & \qty{7}{OV} & \qty{9}{OV} & units & $\nicefrac{\Delta p}{p}$ \\
\hline
Scintillation light yield & 11.0 & 12.3 & 13.1 & $\nicefrac{pe}{keV}$ & 10\% \\
\hline
$\lambda_\textrm{iCT}$ & 18.3 & 29.7 & 41.8 & \% & 3\% \\
\hline
$\lambda_\textrm{eCT}$ & 9.2 & 14.9 & 20.9 & \% & 3\% \\
\hline
$P_\textrm{AP}^{(\nicefrac{1}{2})}$ & 3.2 & 4.6 & 5.9 & \% & 10\%\\
\hline
SNR & 14 & 10 & 7.8 & - & 5\%\\
\hline
\end{tabular}
\caption{Input parameters for the Toy Montecarlo simulation extracted from~\cite{star} at 5, 7, 9 over-voltages: the last column quantifies the relative uncertainty for each parameter at one sigma. The eCT includes the contribution of the feedback cross-talk measured in STAR (see text).
The light yield and the external cross-talk are rescaled by the light losses of the STAR chamber (\qty{13}{\percent}).} 
\label{tab:tmc}
\end{table}

\section{Toy Monte Carlo simulations}
\label{sec:tmc}

To properly account for all noise contributions, a dedicated toy Monte Carlo (tMC) simulation was developed: the code is based on the simulation used in STAR R\&D facility at LNGS~\cite{star}, i.e. a down-scaled liquid argon detector, instrumented with the aforementioned NUV-HD-Cryo SiPMs. The tMC was extended to a large homogeneous liquid argon detector with $N_\textrm{pd}\simeq 2000$  readout channels based on SiPM photo-detectors. 
The SiPM parameters and scintillation light yield  were extracted from STAR, Tab.~\ref{tab:tmc}.

The simulation does not include the scintillation photon tracking (as for a full chain physical Monte Carlo simulation) and the light collection efficiency is considered uniform in the fiducial volume: all photo-detectors have the same probability of being hit by scintillation or cross-talk photons. As a consequence of the simplification, the following results has to be considered a pure and ideal case, as just driven by the statistical features of the SiPM intrinsic noise. The real physical performances will be further reduced by the detector imperfections, as detector reflectivity, scintillation light absorption, TPB re-emission, electronics flaws and all other possible systematic effects.

For each detected photo-electron (from primary argon scintillation or from correlated noises) a recursive oCT+AP cascade is applied. This is implemented by a set of recursive functions each describing one possible correlated noise. Each function generates the correlated noise using a binomial extraction (or two binomial for iCT, as in~\cite{star}). If an avalanche is generated, a recursion on all correlated noises is activated. In case of eCT, the avalanche is applied to another readout channel, randomly picked. All the oCT and AP components are accounted for individually. It is important to notice that the tMC does not depend on the models described in the previous section that describe the average cascade behaviour (Eq.~\ref{eqn:tartaglia} and \ref{eqn:gain:octap}): the tMC propagates each photo-electron independently.

The tMC keeps track of photo-electrons (both primaries and secondaries) generated by the singlet and triplet argon states for electron and nuclear recoils. 
The median values for the fast-to-total light yield (typically referred to as $f_{90}$ or $f_{\rm prompt}$) are extracted from~\cite{bib:SCENE, bib:UAr}.
As discussed above, after-pulsing closer than $\tau_{\nicefrac{1}{2}}^\textrm{AP}$ has smaller amplitude and reduced probability of triggering oCTs. 
It is safe to assume that the after-pulsing (and its secondaries) does not contribute to the detected singlet photo-electrons with a timing of few nanoseconds. Notice that, for large experiments, the propagation time of the photons in the liquid argon chamber can increase the spread of the singlet photo-electrons beyond $\tau_{\nicefrac{1}{2}}^\textrm{AP}$. In such case the after-pulsing can affect the singlet too. Since in any case the AP probability is very small for the NUV-HD-Cryo, this case is not particularly interesting.\\

The region of interest (RoI) for the analysis is between is in the range 5--35 keVee: the tMC simulates a wider window to avoid border effects due to fluctuations.

\subsection{Energy reconstruction}
In absence of correlated noises, it is possible to define three estimators for the \textit{net} energy of the scintillation event. $N_N$ is calculated as the total number of avalanches detected by photo-detectors. $B_N$ is the number of read-out channels with at least one avalanche. In first approximation, $B_N$ behaves as binomial distribution with success probability $1-e^{-N_N/N_\textrm{pd}}$ with $N_\textrm{pd}$ extractions (i.e. the total number of read-out channels).
Therefore,
\begin{equation} \label{eq:B}
    \langle B_\textrm{N} \rangle = N_\textrm{pd}(1-e^{-N_N/N_\textrm{pd}})\, ,
\end{equation}
where the mean value is required because the binomial extraction introduces an irreducible spread in the data. 
To overcome the inherent non-linearity of $B_N$ as function of energy, the $L_N$ (linearized binomial) variable is introduced with the following definition:
\begin{equation}  \label{eq:L}
    L_N = -N_\textrm{pd} \ln \left(1-\frac{B_\textrm{N}}{N_\textrm{pd}} \right)\, .
\end{equation}

\begin{figure}
    \centering
    \includegraphics[width=0.6\columnwidth]{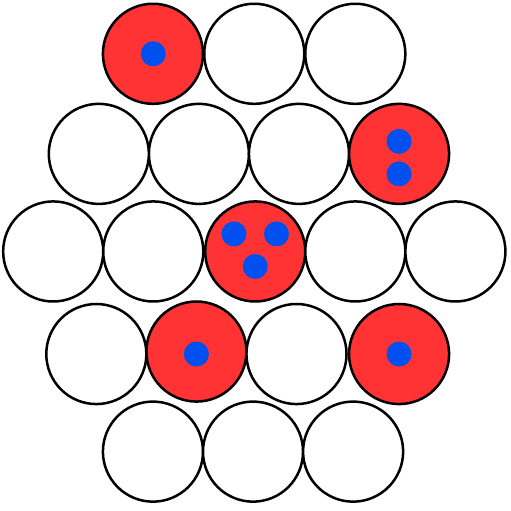}
    \caption{Example of energy estimator during the event reconstruction. For both SiPM or PMTs, the energy of an events producing several photo-electrons can be estimated counting the number of total photo-electrons ($N$, blue circles) or counting the number of total photo-detectors ($B$, red circles).}
    \label{fig:var}
\end{figure}

Figure \ref{fig:var} intuitively depicts the meaning of the binomial counting: for $B_N$ the information of piled-up photo-electrons in the same read-out channel is lost. The information loss leads inevitably to a spoiled resolution, however the fluctuations introduced  by the correlated noise are strongly reduced, as it will be shown later. 
As hinted earlier, the binomial extraction is only an approximation because the number of photo-electron is set to $N_N$ and only the pile-up can fluctuate. This is clearer for an event with $N_N = 1$: the binomial extraction allows non-physical events with $B_N\ne1$.
The tMC correctly simulates the pile-up and the reported results are unaffected by the binomial approximation used to simplify the description.\\

\begin{figure}[tb]
\centering
\includegraphics[width=\columnwidth]{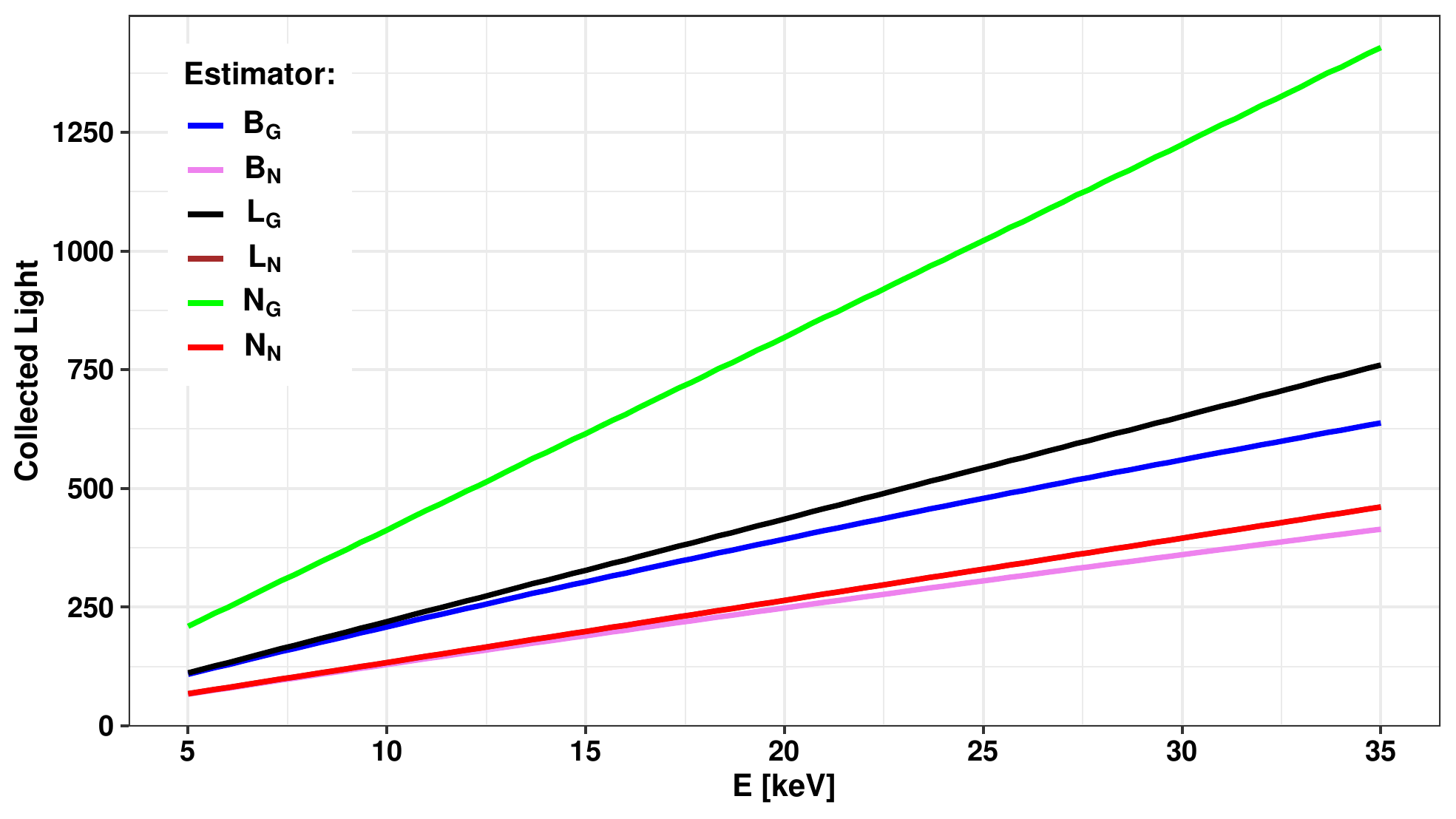}
\caption{Average collected light for electron recoils for the different estimators described in the text. The $L_N$ curve is completely overlapped by the $N_N$ line. With the exception of the $B^*$ variables, the curves are straight line with slope defined by the light yield at \qty{9}{OV} multiplied by the correlated noises gain, Equations~\ref{eqn:gain:octap} and~\ref{eqn:tartaglia}. In the rest of the article $N_N$, $N_G$, $L_N$ and $L_G$ will be used as energy estimators by implicitly inverting this plot.}
\label{fig:energy:scale}
\end{figure}

Similar quantities can be calculated in presence of correlated noises (oCT and AP), obtaining $N_G$, $B_G$, $L_G$ for modelling the \textit{gross} energy seen by the detector.
Figure~\ref{fig:energy:scale} reports the reconstructed values of $N$, $B$ and $L$ (net and gross) as a function of the deposited energy for electron recoil events ($E_e$) in keVee. 
Both $L$ and $N$ provide a linear scale to measure the released energy (once a calibration is performed) and will hence be used as energy estimators. It is interesting to notice that the $L_G$ estimator is less affected by internal cross-talk and after-pulsing than $N_G$, since only the first photo-electron from scintillation (or from eCT) is accounted for by each photo-detector. As a result, the noisy gain  of the correlated processes (the slope of Fig.~\ref{fig:energy:scale}) is unequivocally lower then $N_G$: from Eq.~\ref{eqn:tartaglia}, $\mathcal{G}_{L_G} = 1 + \mathcal{G}_\textrm{oCT+AP}\,\lambda_\textrm{eCT}$. 

\begin{figure}[tb]
\centering
\includegraphics[width=\columnwidth]{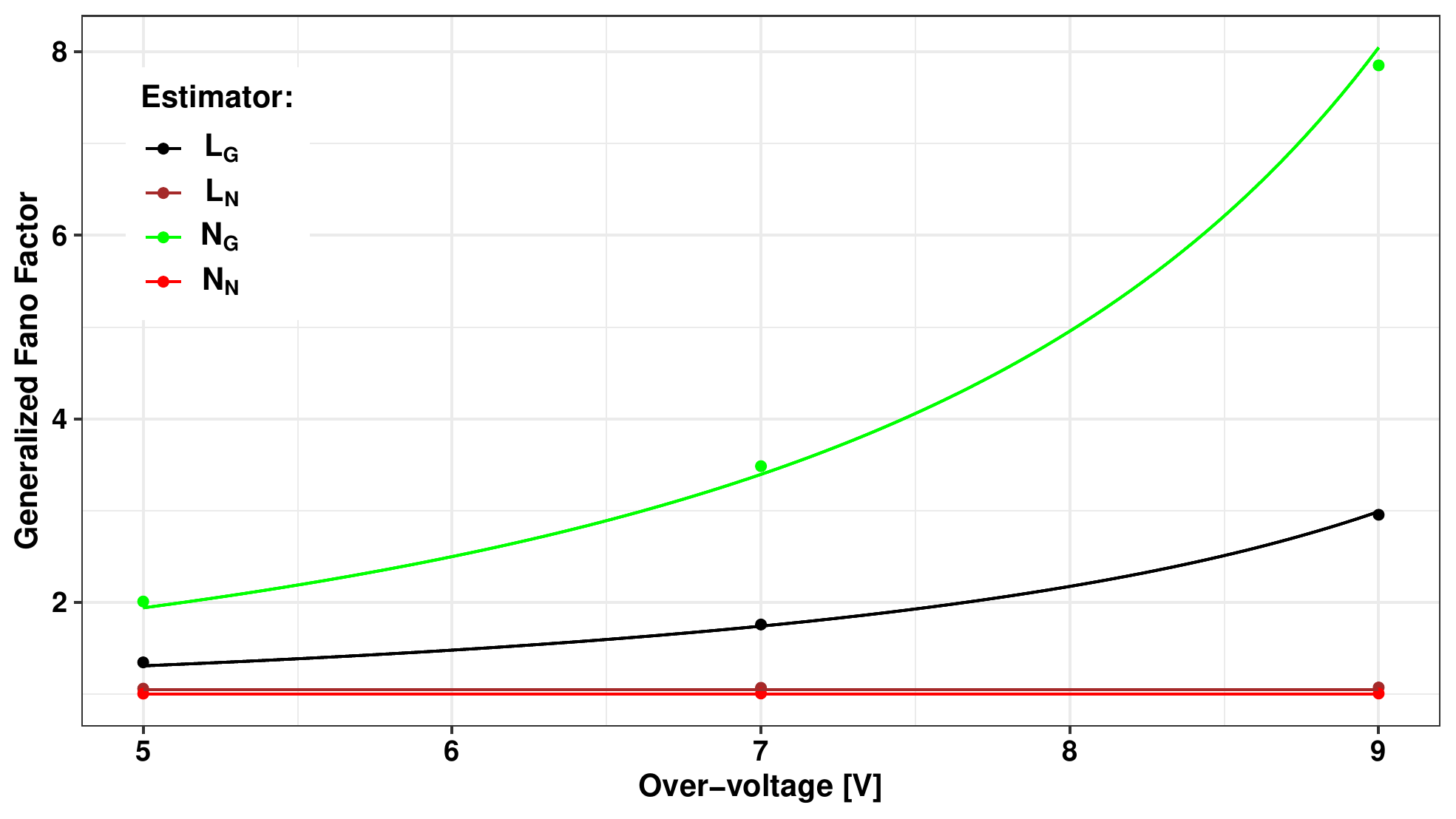}
\caption{Average generalized Fano factor for electron recoil events versus SiPM over-voltages for the energy estimators described in the text. As expected $\mathcal{F}_{N_N}$ has unity value, being the process purely Poissonian. 
The generalized Fano factor for the $L$ estimators ($\mathcal{F}_{L_N}$ and $\mathcal{F}_{L_G}$) is inherently not constant as function of the energy. Since in the region of interest it changes by less than \qty{7}{\percent}, the mean value is used. The lines for $\mathcal{F}_{N_G}$ and $\mathcal{F}_{L_G}$ represent the best-fit described in the text, while for $\mathcal{F}_{N_N}$ and $\mathcal{F}_{L_N}$ a constant line at 1 is drawn to guide the eye.}
\label{fig:energy:gf}
\end{figure}

Figure~\ref{fig:energy:gf} reports the generalized Fano factor (defined as variance over mean of the collected number of photo-electrons) for different energy estimators. 
It is possible to express the energy resolution of the experiment as a function of the correlated noise gains. 
A non-linear regression analysis on the simulated data shows an exponential dependence of the Fano factors upon the the noise gains, namely $\mathcal{F}_{N_G}=(\mathcal{G}_\textrm{oCT+AP})^{-1.84\pm0.02}$ and $\mathcal{F}_{L_G}=(\mathcal{G}_{L_G})^{-2.18\pm0.04}$, with a standard deviation of the relative residuals of $\sim$\qty{3}{\percent}.

The lower generalized Fano factor for $L_G$ with respect to $N_G$, and hence the better energy resolution in the RoI, is directly linked to the lower noise gain of the oCT process affecting the linearized binomial energy estimator. For example, at \qty{9}{OV} $\mathcal{G}_{L_G}\simeq \nicefrac{1}{2}~ \mathcal{G}_\textrm{oCT+AP}$.

\subsection{Pulse shape}
\begin{figure*}[tb]
\centering
\includegraphics[width=.85\textwidth]{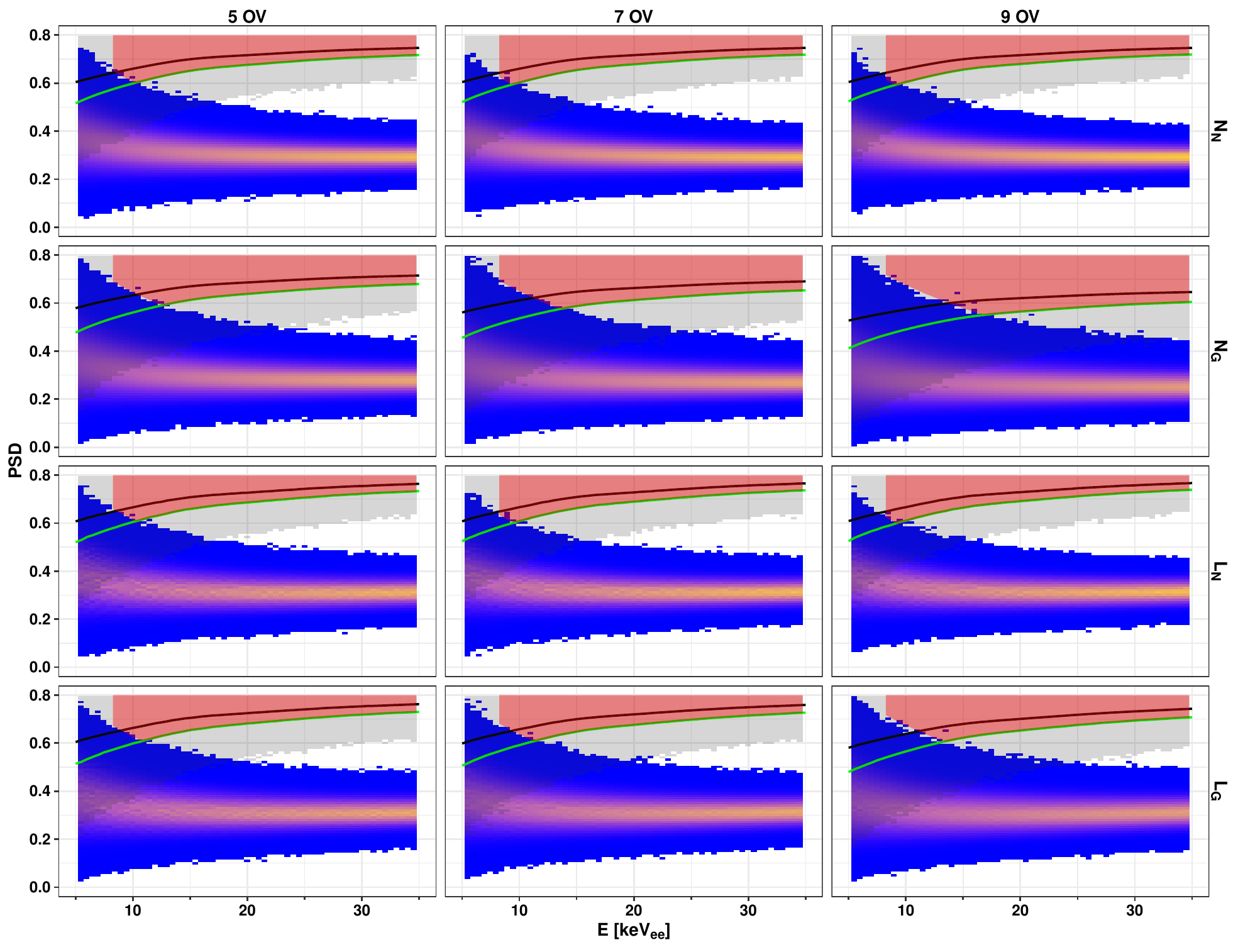}
\caption{Pulse shape discrimination versus reconstructed energy for several SiPM over-voltages and different energy estimators. The blue-gold heat-maps correspond to \num{3E10} electron-like events uniformly distributed between \qtyrange[range-phrase = \ \text{and}\ ]{5}{35}{\keV_{ee}}. The gray shaded bands correspond to about \num{1E9} nuclear recoil events: \qty{50}{\percent} (black) and \qty{90}{\percent} (green) quantiles are shown. The red ribbons identify the dark matter acceptance area with a leakage of about ten electron like events.}
\label{fig:psd}
\end{figure*}

The tMC keeps track of which photon corresponds to the singlet or triplet emission in liquid argon scintillation. The pulse discrimination parameter (PSD) is defined as the ratio between the number of singlet photo-electron collected divided by the total number of photo-electrons.
Figure~\ref{fig:psd} reports the PSD as a function of the reconstructed energy for \num{3E10} electron-like events for the energy estimators discussed earlier and with three SiPM over-voltages, namely 5 V, 7 V and 9 V.
As expected, these plots are affected by the presence of correlated noise with a spread in PSD and by a lower resolution on the energy scale.  

\subsection{S2}
As mentioned before, S2 plays a crucial role in the position and energy reconstruction of the double-phase argon TPCs and is fundamental for the pulse shape of double-phase xenon TPC. Typically, S2 is much greater than S1 and depends on applied drift and extraction fields. For the reasons discussed above, S2 is indeed not free from all of the issues concerning the presence of the oCTs. Furthermore, in S2 the scintillation photo-electrons are not uniformly distributed: only a few photo-detectors, just above the S2 position, will see the largest fraction of the emitted light. In these conditions, the use of binomial estimators is not possible.
The full description of eCT effects on S2 would anyway require a detailed Monte Carlo analysis, e.g. for the position reconstruction algorithm, that goes beyond the scope of the present work.

\section{Implications on the projected sensitivity}
\label{sec:sensitivity}
Assuming the SI WIMP-nucleon interaction, under the (galactic) Standard Halo model, it is possible to extract the projected sensitivity of the simulated experiment.
The number of background events is considered zero: the suppression of nuclear recoil events is demanded to a proper detector design, including 
active neutron veto and low background materials.
For leaking electron recoil this is obtained by tuning the NR acceptance regions. This process depends strongly on the correlated noise contribution and on the used energy estimator as discussed in Sec. \ref{sec:tmc}. 

\subsection{Acceptance Regions}
\begin{figure}[tb]
\centering
\includegraphics[width=\columnwidth]{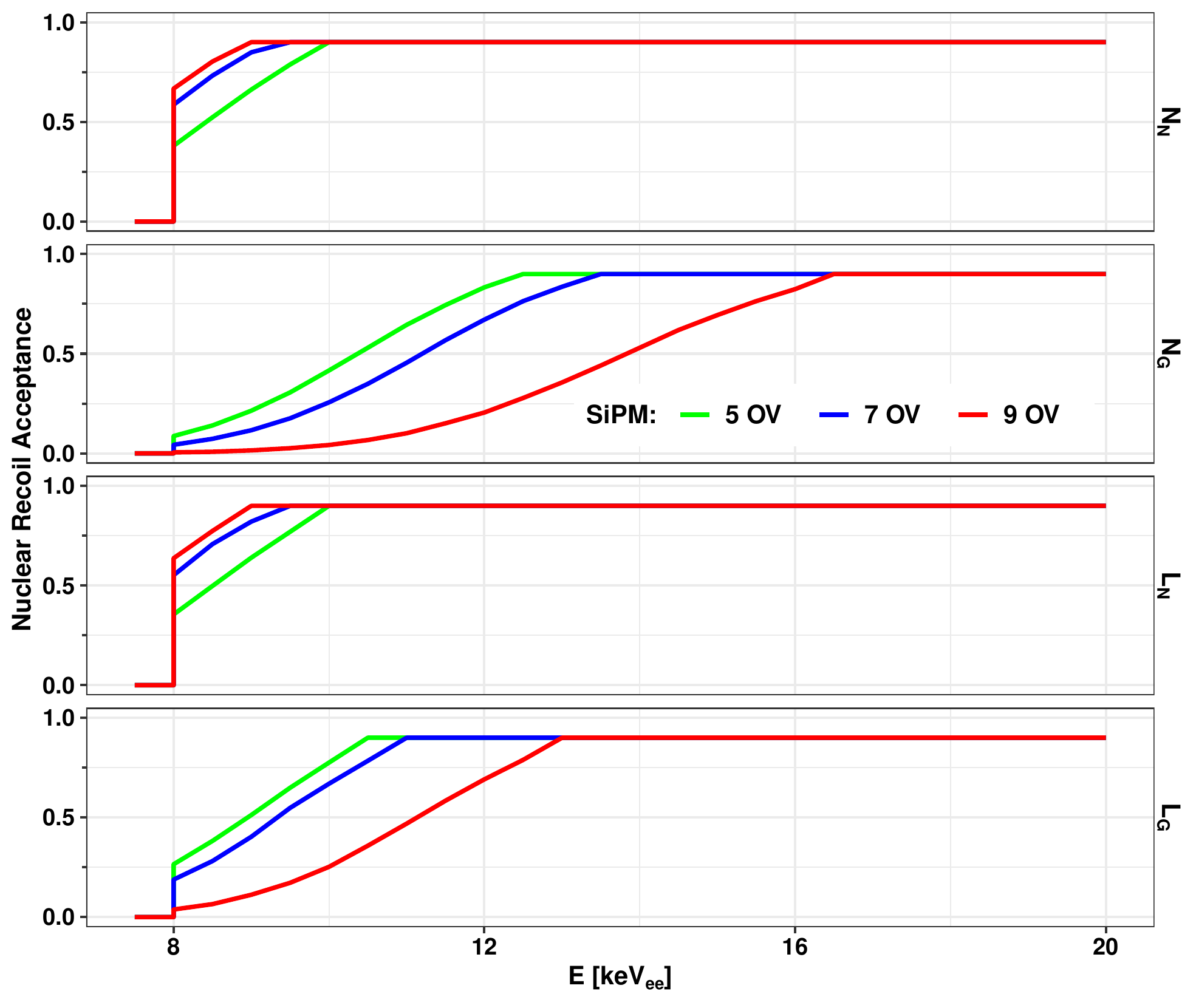}
\caption{Nuclear recoil acceptances as a function of collected energy for the different estimators discussed in the text and at several over-voltages. Above 20 keVee the acceptance remains flat at \qty{90}{\percent} and is not shown and below 8 are set to zero.}
\label{fig:acc}
\end{figure}

Based on the data in Fig.~\ref{fig:psd}, the acceptance regions are defined as the areas in which only about \num{10} ER leaking events are present over the aforementioned exposure. The acceptance regions are defined between \qtyrange[range-phrase = \ \text{and}\ ]{5}{35}{\keV} by the intersection of the \qty{90}{\percent} quantile for NR events and a segment of a rectangular hyperbola ($\nicefrac{1}{(1 + \xi E)}$) in the pulse shape parameter. The scale parameter ($\xi$) is tuned to satisfy the requirement on having \num{10} ER outliers in the acceptance for each SiPM OVs and energy estimators. The resulting NR acceptance as function of energy is shown in Fig.~\ref{fig:acc}.

This number of leaking events corresponds to \num{0.1\pm0.01} events over the 200 ~ton$\cdot$y exposure.
Indeed, neglecting the contribution of all possible $\gamma$ background in the detector fiducial volume, the only source of internal ER events is the $^{39}$Ar $\beta$ decay, and amount to 
\begin{equation}
   N_{\isotope[39]{Ar}} = 3.15 \cdot 10^{10} \times \frac{\rm mass}{\rm [ton]} \times \frac{\rm time}{\rm [y]}\times\frac{\delta _\textrm{RoI}}{\mu}\, ,
\end{equation}
where $\mu$ is the \isotope[39]{Ar} depletion factor, reasonably assumed to be about $1/1500$ \cite{bib:UAr},
with respect to specific activity of the atmospheric argon (1 Bq/Kg);  $\delta_\textrm{RoI}$ is the fraction of the \isotope[39]{Ar} beta decays falling in the RoI (calculated by integrating the normalized $\beta$--spectrum over the RoI). 
For the chosen exposure, one obtains $N_{\isotope[39]{Ar}}\simeq\num{3e8}$ events.

\subsection{Dark matter sensitivity plots}

To calculate the dark matter sensitivity plot, the standard WIMP-halo model with $v_{\rm esc} = 544$ km/s \cite{bib:Smith}, $v_0= 220$ km/s \cite{bib:Smith}, $v_{\rm earth} = 232$ km/s \cite{bib:Savage}, $\rho_{\rm DM} = 0.3$ GeV/(c$^2$ cm$^3$) \cite{bib:Freese} is assumed. 
The SI WIMP-nucleon differential interaction rate, as a function of the kinematic parameters and of the dark matter velocity distribution, is convoluted with the energy resolution for each energy estimator. The number of interactions is given by the exposure multiplied by the differential rate integrated over the RoI, scaled by the acceptance of Fig.~\ref{fig:acc} and by the exposure. The NR quenching, as a function of the energy for liquid argon, is taken from \cite{bib:UAr}.
Assuming the null result and neglecting the 0.1 \isotope[39]{Ar} background events, the \qty{90}{\percent} CL exclusion curves, corresponding to the observation of 2.3 events for the SI WIMP-nucleon interactions, are derived for each energy estimator. 

\begin{figure}[tb]
\centering
\includegraphics[width=\columnwidth]{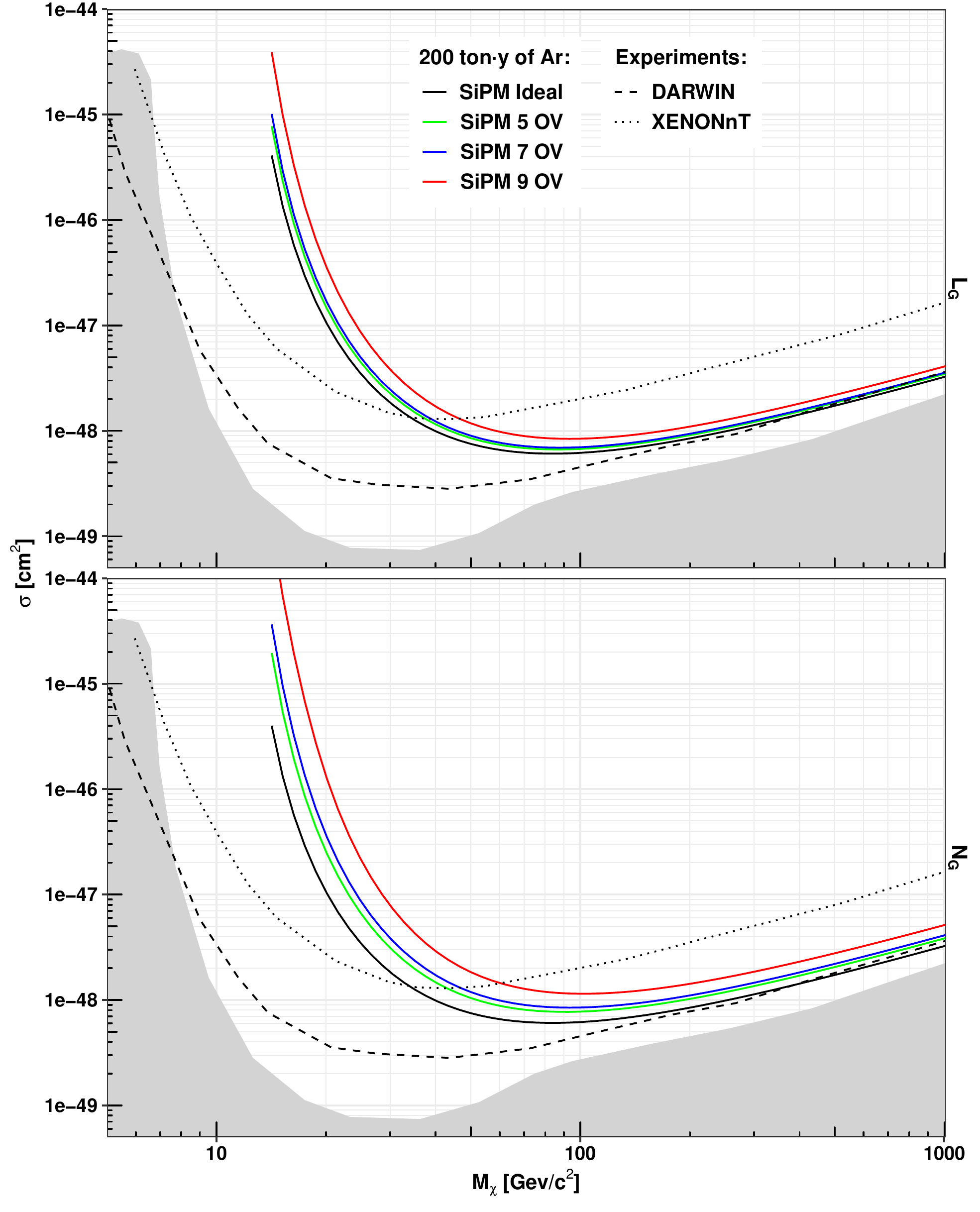}
\caption{Sensitivity curve (90\% CL) for a 200 ton$\cdot$y exposure of a liquid argon detector using SiPM-based photo-detector module for different over-voltages and using two different energy estimators for S1: linearized binomial counting (top) and number of detected photo-electrons (bottom). The ideal curves refers to the absence of noises (oCT, AP and electronic).
As reference the projected sensitivity curves for XENONnT~\cite{xenonnt_sens} and for DARWIN~\cite{darwin} are added with dotted and dashed lines. The neutrino floor~\cite{bib:PDG} with gray area is also shown. As explained in the text, the projected sensitivity for argon detector with SiPMs is overestimated by a factor $2\div3$ with respect to a realistic experiment.}
\label{fig:sens}
\end{figure}

Figure \ref{fig:sens} shows the corresponding sensitivity curves for the energy estimators described earlier for different SiPM over-voltages. The Figures show a progressive decrease of the sensitivity, depending on the operating over-voltage. In other words, the optical correlated noise produces a sizeable effect on the final analysis that may reduce the projected sensitivity even by a factor two, especially in case a large OV is used. This effect can be mildly reduced using the linearized binomial energy estimator, but not completely removed.

Notice that the ideal sensitivity curves are tangibly better than a realistic experimental situation because, as explained in Sec. \ref{sec:tmc}, the tMC does not take into account a further broadening of the energy and the pulse shape resolutions, due to the scintillation properties and the detector geometry. Considering the acceptance of a real liquid argon detector as DarkSide-50 \cite{ds532}, one can easily estimate that sensitivity is overestimated by a factor $2\div3$, or even more.
For this reason, the distortion caused by the correlated noise with respect to the ideal Poisson baseline must be considered in a relative and not in an absolute way, and its real impact could be dramatically large, depending on how large is the difference between the reconstruction of the physical observables  and their ideal expectations.

\section*{Conclusions}
The direct dark matter search, especially concerning the WIMP-like detection in the mass interval 1 GeV -- 1 TeV, starts exploiting the multi-ton scale. For the future detectors, scalable and reliable photo-detectors are required. The argon community decided to use SiPMs taking advantage of the high PDE and low background. SiPMs have been developed at FBK for operation in liquid argon and large photo-detectors have been proven at LNGS, opening the path to the deployment of large experiments.
However, as largely discussed in the present work, the presence of correlated noise can spoil the full scientific reach of future detectors if not properly managed.

We introduced a new energy estimator, based on the number of fired module, which is capable of mitigating the noise caused by optical cross-talk and after-pulsing.
Differently, the external optical cross-talk is irreducible and become dominant for high over-voltages. We have shown, as an example, a multi-ton liquid argon detector with 200 ton$\cdot$y exposure. A dedicated simulation shows that the presence of the irreducible correlated noise can even halve the ideal projected sensitivity (Fig.~\ref{fig:sens}), basically doubling the needed experiment live time. The xenon-based detector should, in principle, be less affected by this issue, as the sensitivity for optical cross-talk of VUV SiPM is lower. 

Beyond analysis optimizations, it is possible, in principle, to mitigate in the detectors the effects of correlated noise. A first solution could be the introduction of colored optical filters capable of attenuating wavelengths above \qty{500}{\nano\meter} in front of the SiPM-based modules.
More elaborate options require modifications to the SiPM to reduce the iCT, the emission probability and the PDE above the green wavelength.

As we have shown in our toy Monte Carlo simulations, the correlated noise largely applies for high over-voltage settings, where the effect is strongly amplified. Therefore, a natural solution could be operating the detectors at very low over-voltage, to reduce the cascade gain down to some acceptable value. However, reducing the SiPM charge gain, may required a very low electronic noise condition that can not be easily achieved in very big detectors.

The dark matter search is entering a critical and challenging phase, in which old technological solutions may not be sufficient any longer and novel and promising solutions has to be carefully tested and validated in dedicated prototypes. And SiMPs detectors are a neat example in this sense.

\section*{Acknowledgements}
We'd like to thank the LNGS computing center and Dr.~C.~Pellegrino at CNAF for providing the CPU required to run the simulations. We acknowledge the support of Dr.~G. Rannucci and Prof.~V.~Caracciolo, in the discussion of the statistical models and for the proofreading.

\bibliographystyle{spphys}
\bibliography{main, nicola}

\end{document}